\documentclass[10pt,letterpaper]{article}
\usepackage[margin=1in,nohead,centering]{geometry}
\usepackage{setspace}
\newcommand{\mathsym}[1]{{}}
\newcommand{\unicode}[1]{{}}

\usepackage{graphicx,color,wrapfig}
\usepackage{amssymb}
\usepackage{amsmath}
\usepackage{url}
\usepackage{xspace}
\usepackage{float}
\usepackage{subfigure}
\usepackage{graphics}

\newcommand{\aseq}{{\bf a}\xspace}

\newtheorem{theorem}{Theorem}
\newtheorem{example}{Example}

\newtheorem{lemma}[example]{Lemma}

\chardef\other=12
\def\mdeactivate{%
\catcode`\&=\other   \catcode`\#=\other
\catcode`\%=\other   \catcode`\~=\other
}
\def\mmakeactive#1{\catcode`#1=\active\ignorespaces}

{
\mmakeactive\
\gdef\obeywhitespace{%
  \mmakeactive\^^M %
  \let^^M=\NewLine %
  \aftergroup\removebox %
  \obeyspaces %
}}

\def\NewLine{\par\indent}
\def\removebox{\setbox0=\lastbox}

\def\mverbatim{\par\begingroup\parindent=0em\tt\mdeactivate\obeywhitespace
\catcode`\|=0  %
}

\def\|{|}

\title{Asymptotic connectivity for the network of RNA secondary structures}
\author{P. Clote}
\date{Biology Department, Boston College, Chestnut Hill, MA 02467,
{\tt clote@bc.edu}}

\begin{document}

\maketitle

\begin{abstract}
Given an RNA sequence $\aseq$, consider the network $G=(V,E)$, where the set
$V$ of nodes consists of all secondary structures of $\aseq$, and whose
edge set $E$ consists of all edges connecting two secondary structures
whose base pair distance is $1$. Define the network {\em connectivity},
or {\em expected network degree}, as the average number of edges incident
to vertices of $G$.  Using algebraic combinatorial methods,
we prove that the asymptotic connectivity of length $n$ homopolymer
sequences is $0.473418 \cdot n$. This raises the question of what other
network properties are characteristic of the network of RNA secondary 
structures. Programs in Python, C and Mathematica are available at the
web site \url{http://bioinformatics.bc.edu/clotelab/RNAexpNumNbors}.
\end{abstract}

\section{Introduction}

In \cite{steinWaterman}, Stein and Waterman used generating function theory
to determine the asymptotic number of RNA secondary structures.
Since that pioneering paper, a number of results concerning RNA
secondary structure asymptotics have appeared, including the incomplete list
\cite{Hofacker1998Combinatorics,Haslinger-Stadler-1999,Nebel.jcb02,Clote.jcb06,Lorenz.jcb08,Clote.jbcb09,Han2009Stacks,Jin2010Decomposition,Reidys2010Modular,Reidys.b11,Clote.jmb11,Saule.jcb11,Han.jcb12,Li.jmb12,Fusy.jmb12}.

In contrast to the previous papers, here we consider 
{\em network} properties of the ensemble of RNA secondary structures,
following the seminal work of Wuchty
\cite{Wuchty.nar03}, who showed by exhaustive enumeration of the low energy
secondary structures of {\em E. coli} phe-tRNA, that the corresponding
network architecture had {\em small-world} properties \cite{Watts.n98}.
Small-world networks appear to abound in biology, providing a kind of
robustness necessary for the molecular processes of life, as seen in 
networks of neural connections of {\em C. elegans} \cite{Watts.n98},
gene co-expression in {\em S. cerevisiae} \cite{VanNoort.er04},
metabolic pathways \cite{Wagner.pbs01,Ravasz.s02},
intermediate conformations in tertiary folding kinetics 
for the protein villin \cite{Bowman.pnas10}, etc. 

In this paper, we use algebraic combinatorial methods, and in particular
the Flajolet-Odlyzko theorem \cite{FlaOdl90} to prove that the asymptotic
expected network connectivity of RNA secondary structures is
$0.4734176431521986 \cdot n$.  Following \cite{waterman,hofacker99a},
{\em stickiness} is defined to be the probability $p$
that any two positions can pair. For the simplicity of argument, in the
homopolymer model, we take stickiness $p$ to be $1$; however, minor
changes in our C dynamic programming algorithm and in our Mathematica code
permit the computation of asymptotic expected connectivity for arbitrary
stickiness $p$.

\section{Preliminaries}

A secondary structure for a given RNA nucleotide sequence 
$\aseq = \aseq_1,\dots,\aseq_n$ is a set $s$ of base pairs $(i,j)$, such that
{\em (i)} if $(i,j)\in s$ then 
$\aseq,\aseq_j$ form either a Watson-Crick (AU,UA,CG,GC) or 
wobble (GU,UG) base pair, 
{\em (ii)} if $(i,j)\in s$ then $j-i>\theta=3$ (a steric constraint
requiring that there be at least $\theta=3$ unpaired bases between
any two paired bases),
{\em (iii)} if $(i,j)\in s$ then for all $j' \ne j$ and $i' \ne i$,
$(i',j) \not\in s$ and $(i,j') \not\in s$ (nonexistence of base triples),
{\em (iv)} if $(i,j)\in s$ and $(k,\ell)\in s$, then 
it is not the case that $i<k<j<\ell$ (nonexistence of pseudoknots).
For the purposes of this paper, following Stein and 
Waterman \cite{steinWaterman}, we consider the {\em homopolymer}
model of RNA, in which condition 
{\em (i)} is dropped, so that any base can pair with
any other base.

Suppose that $\aseq=\aseq_1,\ldots,\aseq_n$ is an RNA sequence.
If $s$ is a secondary structure of $\aseq$, then let $N_s$ denote the number of
secondary structures of $\aseq$ that can be obtained from $s$ by the removal or 
addition of a single base pair; i.e. those structures having base pair
distance from $s$ of $1$.
Define the {\em expected number of neighbors} $\langle N_s \rangle$ by
\begin{eqnarray}
\label{eqn:expectationNx1}
\langle N_s \rangle &=& \frac{Q}{Z}
\end{eqnarray}
where $Q = \sum_s N_s$ is the total number $N_x$ of neighbors of all
secondary structures $s$ of $\aseq$,
and $Z$ denotes the total number of secondary
structures of $\aseq$.
Note that $Z$ corresponds to the {\em partition function}
$\sum_s \exp(-E(s)/RT)$ if the energy of every structure is $0$.

In \cite{cloteJCC2014} we described three algorithms to compute
the expected number of neighbors, or {\em network connectivity}, 
$\langle N_s \rangle = \sum_{s} \frac{\exp(-E(s)/RT)}{Z} \cdot N_s$,
where $Z$ is the partition function $\sum_s \exp(-E(s)/RT)$
with respect to energy
model A (each structure $s$ has energy $E(s)=0$), 
model B (each structure $s$ has Nussinov \cite{nussinovJacobson}
energy $E(s)$ equal to $-1$ times the number $|s|$ of base pairs in $s$),
and model C (each structure $s$ has energy $E(s)$ given by the
Turner energy model \cite{turner,xia:RNA}).

Below, we follow reference \cite{cloteJCC2014} in deriving the recurrence
relations for $Q$ and $Z$ for model A, corresponding to equation
equation~(\ref{eqn:expectationNx1}).  For $1 \leq i \leq j \leq n$,
define the subsequence
$\aseq[i,j] = \aseq_i,\ldots,\aseq_j$, and define $SS(\aseq[i,j])$ to
be the collection of secondary structures of $\aseq[i,j]$. Define
\begin{eqnarray}
\label{eqn:Qij_baseCase1}
Q_{i,j} = \sum_{s \in SS(\aseq[i,j])} N_s.
\end{eqnarray}
Similarly, let $Z_{i,j} = \sum_{s \in SS(\aseq[i,j])} 1$; i.e. $Z_{i,j}$ denotes
the number of secondary structures of $\aseq[i,j]$.
\medskip

\noindent
{\sc Base Case:}
For $j-i \in \{ 0,1,2,3\}$, $Q_{i,j}=0$ and $Z_{i,j}=1$.
\medskip

\noindent
{\sc Inductive Case:} Let $BP(i,j,\aseq)$ be a boolean function, taking
the value $1$ if positions $i,j$ can form a base pair for sequence $\aseq$,
and otherwise taking the value $0$.  Assume that $j-i > 3$.
\medskip

\noindent
{\sc Subcase A:} Consider all secondary structures $s \in \aseq[i,j]$,
for which $j$ is unpaired. For each structure $s$ in this subcase,
the number $N_s$ of neighbors of $s$ is constituted from the number of
structures obtained from $s$ by removal of a single base pair, together
with the number of structures obtained from $s$ by addition of a single
base pair. If the base pair added does not involve terminal position $j$,
then total contribution to $\sum_{s \in SS(\aseq[i,j])} N_s$ is $Q_{i,j-1}$.
It remains to count the contribution due to neighbors $t$ of $s$,
obtained from $s \in SS(\aseq[i,j])$ by adding the base pair $(k,j)$.
This contribution is given by
$\sum_{k=i}^{j-4} BP(k,j,\aseq) \cdot Z_{i,k-1} \cdot Z_{k+1,j-1}$,
where $Z_{i,i-1}$ is defined to be $1$.
Thus the total contribution to $Q_{i,j}$ from this subcase is
\[
Q_{i,j-1} + 
\sum_{k=i}^{j-4} BP(k,j,\aseq) \cdot Z_{i,k-1} \cdot Z_{k+1,j-1}.
\]
\medskip

\noindent
{\sc Subcase B:} Consider all secondary structures $s \in \aseq[i,j]$
that contain the base pair $(k,j)$ for some $k \in \{i,\ldots,j-4\}$.
For secondary structure $s$ in this subcase, the number $N_s$ of neighbors
of $s$ is constituted from the number of structures obtained by removing
base pair $(k,j)$ together with a contribution obtained by adding/removing
a single base pair either to the region $[i,k-1]$ or to the region
$[k+1,j-1]$.
Setting $Q_{i,i-1}$ to be $0$, these contributions are given by 
\[
\sum_{k=i}^{j-4} BP(k,j,\aseq) \cdot \left[
Z_{i,k-1} \cdot Z_{k+1,j-1} +
Q_{i,k-1} \cdot Z_{k+1,j-1} +
Z_{i,k-1} \cdot Q_{k+1,j-1} \right].
\]
In the current subcase, the contribution to $Z_{i,j}$ is
$\sum_{k=i}^{j-4} BP(k,j,\aseq) \cdot Z_{i,k-1} \cdot Z_{k+1,j-1}$.

Finally, taking the contributions from both subcases together, it follows
that
\begin{eqnarray}
\label{eqn:Qij_inductiveCase1}
Q_{i,j} &=& Q_{i,j-1} + 
\sum_{k=i}^{j-4} BP(k,j,\aseq) \cdot \left[ 2 \cdot Z_{i,k-1} \cdot Z_{k+1,j-1} +
Q_{i,k-1} \cdot Z_{k+1,j-1} +
Z_{i,k-1} \cdot Q_{k+1,j-1} \right]\\
\label{eqn:Nij_inductiveCase1}
Z_{i,j} &=& Z_{i,j-1} + 
\sum_{k=i}^{j-4} BP(k,j,\aseq) \cdot Z_{i,k-1} \cdot Z_{k+1,j-1}.
\end{eqnarray}
It follows that the 
expected number $\langle N_s \rangle$ of neighbors $N_s$ of
structures $s$ of $\aseq$ is $\frac{Q_{1,n}}{Z_{1,n}}$.
Note that the recursion for $Z_{i,j}$ is well-known and
due originally to Waterman \cite{waterman:SecStr}.

To provide concrete intuition for the problem we consider, in 
Figure~1, we present the
list of secondary structures for a homopolymer of length $7$, depicted as
a network having expected connectivity of $2$.
In the left panel of
Figure~2,
we present a histogram for the network connectivity (graph degree or 
number of neighbors), by 
analyzing an exhaustively produced list of all 106,633 structures of
the 20-mer homopolymer. In the right panel of
Figure~2,
we present a plot of the {\em normalized} expected number of neighbors of 
for homopolymers of length 1 to 1000 nt, obtained by
dividing the expected number of neighbors by sequence length.
Clearly there appears to be an asymptotic value for the 
length-normalized expected connectivity, suggesting that it may be
possible to formally prove the existence of this asymptotic value,
a task to which the remainder of the paper is dedicated.

\begin{figure*}
\centering
\begin{minipage}{0.45\textwidth}
\mverbatim
a  .......  6
b  (...)..  1
c  (....).  1
d  .(...).  2
e  (.....)  2
f  .(....)  1
g  ..(...)  1
h  ((...))  2
|mendverbatim
\end{minipage}
\hskip 1cm
\begin{minipage}{0.45\textwidth}
\fbox{
\includegraphics[width=0.4\textwidth]{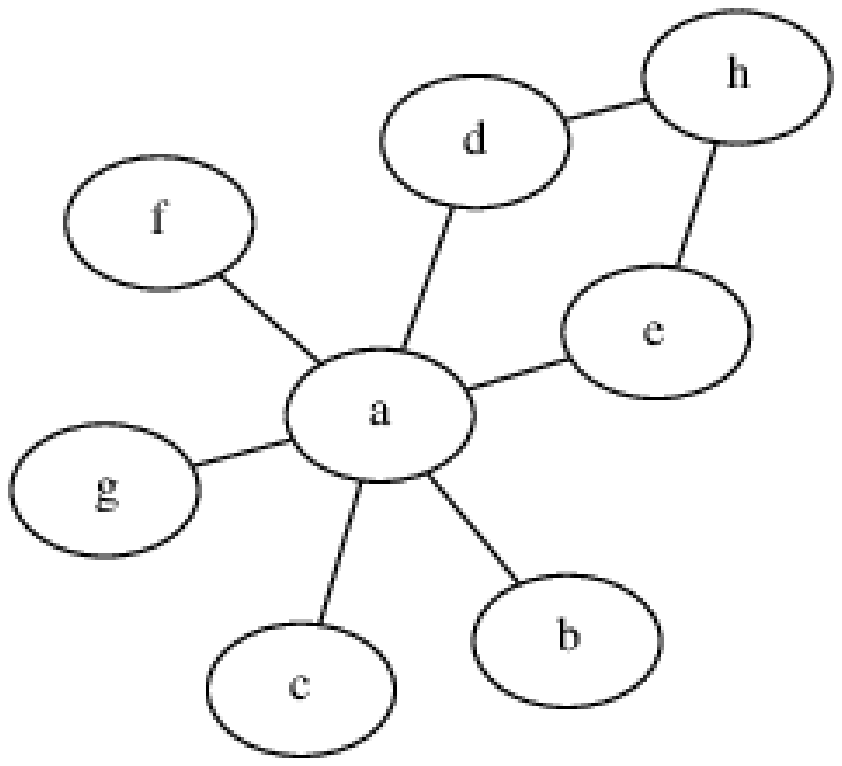}
}
\end{minipage}
\caption{{\em (Left)} All possible secondary structures of the 7-mer
homopolymer, where position $i$ can pair with position $j$ provided only
that $1 \leq i < j \leq 7$ and $j-i \geq 4$.
{\em (Right)} Graph representation of neighborhood network, where
nodes a,b,c,d,e,f,g,h respectively represent the 8 secondary structures
in the list. The number of neighbors of each secondary structure is
indicated to its right. The expected number of neighbors for the 7-mer
is thus $(6+1+1+2+2+1+1+2)/8 = 16/8=2$.
}
\label{fig:toyComputationExpNumNborsAndNeatoGraph}
\end{figure*}

\begin{figure}
\centering
\includegraphics[width=0.45\textwidth]{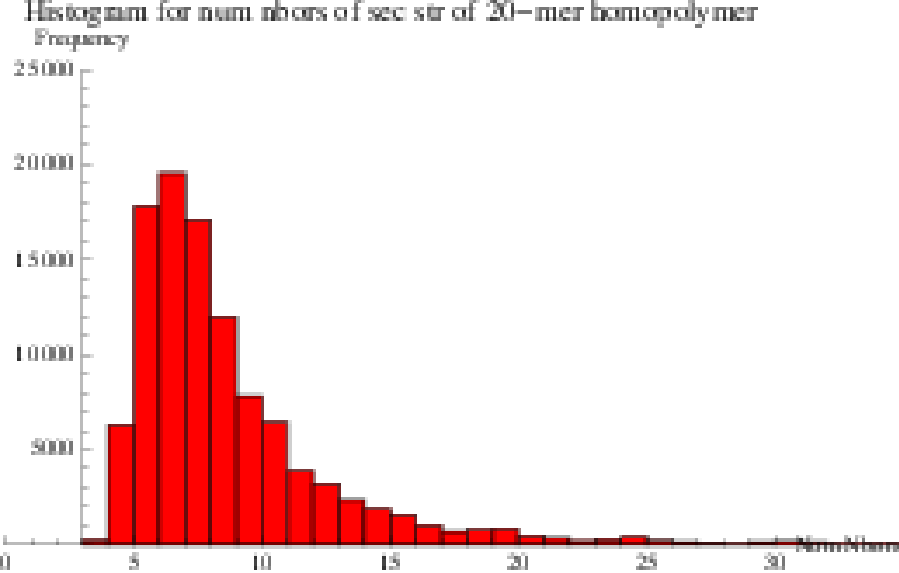}
\includegraphics[width=0.45\textwidth]{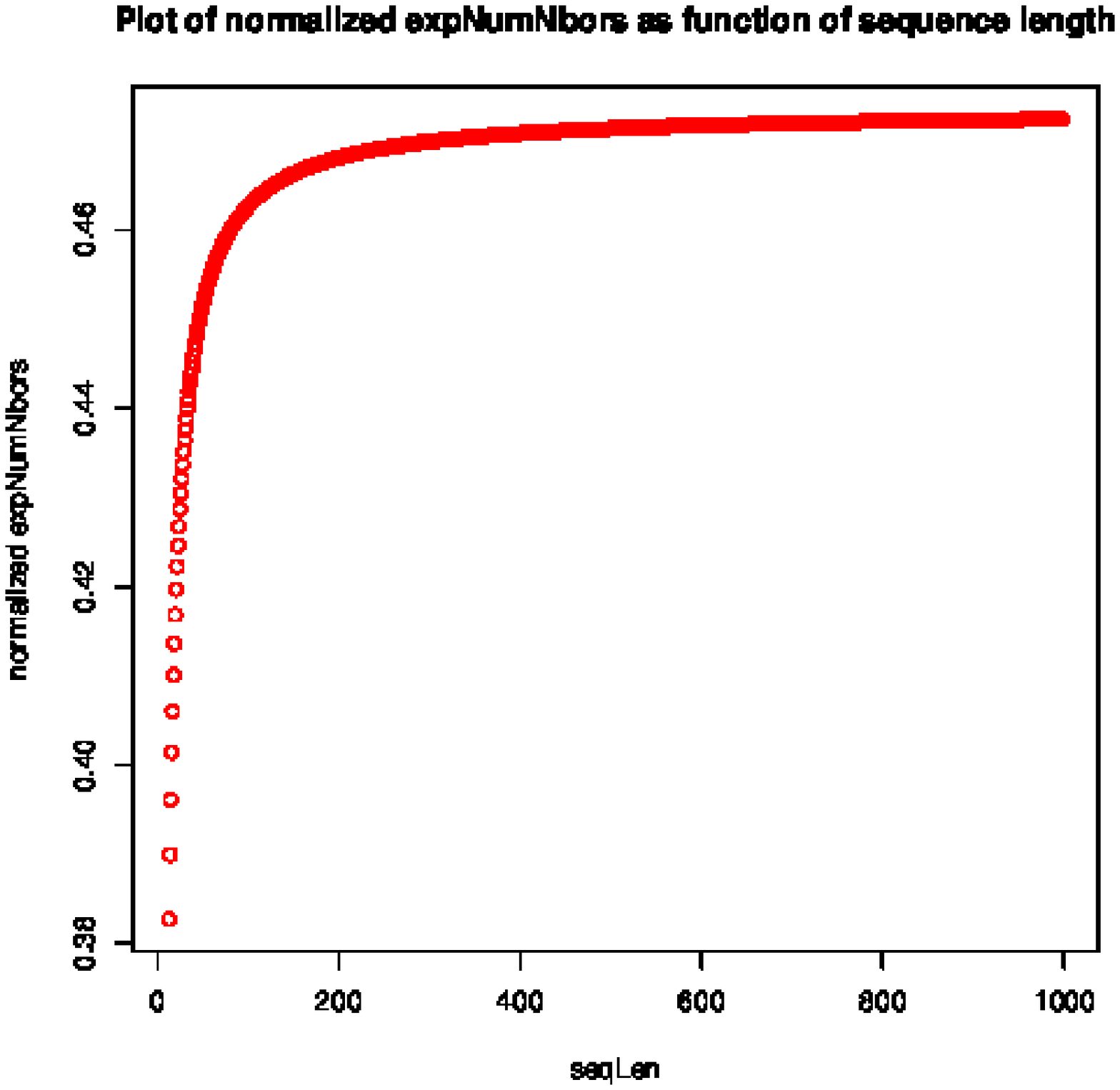}
\caption{{\em (Left)}
Histogram of the number of neighbors for all 106,633 
secondary structures of the  20-mer homopolymer.
The mean is 8.336, the standard deviation is 4.769, the
maximum is 136, and minimum is 3.
{\em (Right)}
Plot of the {\em normalized} expected number of neighbors of 
for homopolymers of length 1 to 1000 nt, obtained by
dividing the expected number of neighbors by sequence length.
Apparent asymptotic value seems to be $\approx 0.4724$. The main
result of this paper is the proof that the asymptotic value is in
fact $0.4734176431521986$.
}
\label{fig:asymptoticNormalizedExpNumNborsHomopolymer}
\end{figure}

\section{Methods}
\subsection{Recurrence relation for partition function $Z_n$}

In order to determine asymptotic network connectivity, 
we now provide similar recursions
to those of (\ref{eqn:Qij_inductiveCase1})
and (\ref{eqn:Nij_inductiveCase1})
for the {\em homopolymer} model of RNA, 
where any base (position) $i$ 
can pair with any other position (base) $j$, provide only that 
$1 \leq i+\theta+1 \leq j  \leq n$. Following common convention due
to steric constraints, we take $\theta$ to be $3$. These recursions
are the basis of the dynamic programming code we implemented, used
to produce Figure~\ref{fig:asymptoticNormalizedExpNumNborsHomopolymer}.

We define $Z_0=1$, in
order to simplify the recurrence relation for $Z_n$, defined to be the
number of secondary structures for a homopolymer of length $n$, 
or equivalently, the {\em partition function} for the energy model that 
assigns an energy of $0$ to every structure.
Moreover, since the empty structure is the only structure for 
sequences of length $1,2,3,4=\theta+1$, we define $Z_n=1$ for 
$0 \leq n \leq 4$. Secondary structures for a homopolymer of length
$n>4$ can be partitioned into two classes: (1) $n$ is unpaired, 
(2) there is a base pair $(k,n)$ for some $1 \leq k \leq n-4 = n-\theta-1$.
Thus we have
\begin{eqnarray}
\label{eqn:partitionFunZn}
Z_n = \left\{ \begin{array}{ll}
1 &\mbox{if $0 \leq n \leq 4 = \theta+1$}\\
Z_{n-1}+\sum_{k=0}^{n-\theta-2} Z_k \cdot Z_{n-2-k} 
 &\mbox{if $n \geq 5 = \theta+2$}
\end{array} \right.
\end{eqnarray}
which is the homopolymer analogue of equation~(\ref{eqn:Nij_inductiveCase1}).
To employ generating function theory, we require a single formula for
$Z_n$, rather than a definition by cases -- see p. 66 of 
\cite{grahamKnuthPatashnik}.  This is easily achieved by {\em (i)} adding the 
indicator function $I[n=0]$, defined to equal $1$ if $n=0$, and otherwise
$0$, and {\em (ii)} adding and subtracting the same terms to ensure that
$k$ ranges from $0$ to $n-2$, rather than $n-\theta-2=n-5$.
Thus equation~(\ref{eqn:partitionFunZn}) is equivalent to
(\ref{eqn:partitionFunZn1}) defined by:
\begin{eqnarray}
\label{eqn:partitionFunZn1}
\nonumber
Z_n &=& Z_{n-1}+\sum_{k=0}^{n-2} Z_k \cdot Z_{n-2-k} 
+I[n=0] - Z_{n-1}\cdot Z_{0} -  Z_{n-3}\cdot Z_1 - Z_{n-4} \cdot Z_2\\
&=& Z_{n-1}+\sum_{k=0}^{n-2} Z_k \cdot Z_{n-2-k} 
+I[n=0] - Z_{n-1} -  Z_{n-3} - Z_{n-4}. 
\end{eqnarray}
Let $z = \sum_{n=0}^{\infty} Z_n \cdot x^n$. By multiplying
equation~(\ref{eqn:partitionFunZn1}) by $x^n$, summing from
$n=0$ to $\infty$, we obtain
\begin{eqnarray*}
\nonumber
\sum_{n=0}^{\infty} Z_n \cdot x^n =
\sum_{n=0}^{\infty} Z_{n-1} \cdot x^n +
\sum_{n=0}^{\infty} \sum_{k=0}^{n-2} Z_k \cdot Z_{n-2-k}  \cdot x^n
- \sum_{n=0}^{\infty} \left(   Z_{n-2} +  Z_{n-3} + Z_{n-4} - I[n=0] \right)
\cdot x^n 
\end{eqnarray*}
which yields
\begin{eqnarray}
\label{eqn:mathematicaZ}
z &=& xz + x^2 z^2 - zx^2 -zx^3 -zx^4 +1
\end{eqnarray}
Solving the quadratic equation~(\ref{eqn:mathematicaZ})
for $z$, we determine that
\[
z=\frac{1-x+x^2+x^3+x^4 \pm \sqrt{1-2 x-x^2+x^4+3 x^6+2 x^7+x^8}}{2 x^2}.
\]
Only the first solution
\begin{eqnarray}
\label{eqn:HauptGleichungZ1}
z1=\frac{1-x+x^2+x^3+x^4-\sqrt{1-2 x-x^2+x^4+3 x^6+2 x^7+x^8}}{2 x^2}
\end{eqnarray}
has the property that the coefficients of its Taylor expansion
correspond to the values of
$Z_n$, as determined by dynamic programming (see program at web site). 
In particular, using Mathematica, we obtain
\begin{eqnarray*}
\label{eqn:mathematicaZseries}
z1 &=& 1+x+x^2+x^3+x^4+2 x^5+4 x^6+8 x^7+16 x^8+32 x^9+65 x^{10}+133 x^{11}+274 x^{12}+568 x^{13}+\\
&& 1184 x^{14}+2481 x^{15}+5223 x^{16}+11042 x^{17}+23434
x^{18}+49908 x^{19}+106633 x^{20}+O(x)^{21}
\end{eqnarray*}
which can be compared with the values determined by our dynamic programming
implementation:
\medskip

\noindent
\begin{tabular}{|l|rrrrrrrrrrrrrrrrrr|}
\hline
$n$&0&1&2&3&4&5&6&7&8&9&10&11&12&13&14&15&16&17\\
\hline
$Z_n$&1&1&1&1&1&2&4&8&16&32&65&133&274&568&1184&2481&5223&11042\\
\hline
\end{tabular}
\medskip

\subsection{ Recurrence relation for the number of neighbors $Q_n$}

Let $Q_n$ denote the total number $N_s$ of nearest neighbors of all
secondary structures $s$ of a homopolymer of length $n$. The
homopolymer analogue of equation~(\ref{eqn:Qij_inductiveCase1}) is
as follows:
\begin{eqnarray}
\label{eqn:homopolymerQn}
Q_n &=&\left\{ \begin{array}{ll}
0 &\mbox{if $n=0,1,2,3,4$}\\
Q_{n-1} + \sum_{k=0}^{n-\theta-2} 2 Z_k Z_{n-2-k} + Q_k Z_{n-2-k} 
+ Z_k Q_{n-2-k} &\mbox{else.}\\
\end{array} \right.
\end{eqnarray}
If we assume that $Z_n=0=Q_n$ for $n<0$, then it follows from
equations~(\ref{eqn:homopolymerQn}) and (\ref{eqn:partitionFunZn}), that
we can express $Q_n$ by the formula
\begin{eqnarray}
\label{eqn:formulaQn}
Q_n = Q_{n-1} + \sum_{k=0}^{n-5} 2 Z_k Z_{n-2-k} + Q_k Z_{n-2-k} 
+ Z_k Q_{n-2-k}.
\end{eqnarray}
Next, we add and subtract the same terms in order to ensure that the
upper bound in the previous summation is $n-2$. This yields
\begin{eqnarray}
\label{eqn:formulaQ1}
Q_n &=& Q_{n-1} + \sum_{k=0}^{n-2} 2 Z_k Z_{n-2-k} + Q_k Z_{n-2-k} 
+ Z_k Q_{n-2-k} \\
\nonumber
&&-2 \left( Z_{n-2} \cdot Z_0 + Z_{n-3} \cdot Z_1 + Z_{n-4} \cdot Z_2 \right)
- \left( Q_{n-2} \cdot Z_0 + Q_{n-3} \cdot Z_1 + Q_{n-4} \cdot Z_2 \right) \\
\nonumber
&&- \left( Z_{n-2} \cdot Q_0 + Z_{n-3} \cdot Q_1 + Z_{n-4} \cdot Q_2 \right)
\end{eqnarray}
which simplifies to
\begin{eqnarray}
\label{eqn:formulaQ2}
Q_n &=& Q_{n-1} + \sum_{k=0}^{n-2} 2 Z_k Z_{n-2-k} + Q_k Z_{n-2-k} 
+ Z_k Q_{n-2-k} \\
\nonumber
&&-2 \left( Z_{n-2} + Z_{n-3} + Z_{n-4} \right)
- \left( Q_{n-2}  + Q_{n-3} + Q_{n-4} \right). 
\end{eqnarray}
Multiply each term of equation (\ref{eqn:formulaQ2}) by $x^n$ and
summing from $n=0$ to $\infty$, to obtain
\begin{eqnarray}
\label{eqn:formulaQ3}
\sum_{n=0}^{\infty} Q_n \cdot x^n = \sum_{n=0}^{\infty} Q_{n-1} \cdot x^n  +
\sum_{n=0}^{\infty} \left( \sum_{k=0}^{n-2} 2 Z_k Z_{n-2-k} \right) + 
\sum_{n=0}^{\infty} \left( \sum_{k=0}^{n-2} Q_k Z_{n-2-k}  \right) + \\
\sum_{n=0}^{\infty} \left( \sum_{k=0}^{n-2}  Z_k Q_{n-2-k} \right) - 
\sum_{n=0}^{\infty}  2 \left( Z_{n-2}+Z_{n-3}+Z_{n-4} \right) - 
\sum_{n=0}^{\infty}  \left( Q_{n-2}+Q_{n-3}+Q_{n-4} \right) .
\nonumber
\end{eqnarray}
Let $q= \sum_{n=0}^{\infty} Q_n x^n$ and $z= \sum_{n=0}^{\infty} Z_n x^n$.
Then from (\ref{eqn:formulaQ3}) we have
\begin{eqnarray}
\label{eqn:formulaQ4}
q=xq+2x^2 z^2 + x^2 qz + x^2 z q -2x^2 z -2x^3 z -2x^4 z -qx^2 -qx^3 -qx^4.
\end{eqnarray}
From equations (\ref{eqn:mathematicaZ}) and (\ref{eqn:formulaQ4}),  we have
\begin{itemize}
\item
$z^2 x^2=z-z x +z x^2 + z x^3 + z x^4 -1$
\item
$q = x q +2x^2 z^2 + 2x^2 q z -2x^2 z - 2 x^3 z -2 x^4 z -q x^2 - q x^3 -q x^4$
\end{itemize}
from which we eliminate the variable $z$ to obtain the following
quadratic equation in variable $q$,
\begin{eqnarray}
\label{eqn:formulaQ4a}
4x^5 &=& q^2 x^2 \left(1-2 x-x^2+x^4+3 x^6+2 x^7+x^8\right)+ \\
&& q\left(2-6 x+2 x^2+2 x^3+2 x^4-2 x^5+6 x^6-2 x^7-2 x^8-2 x^9\right).
\nonumber
\end{eqnarray}
Solving for $q$, we  determine that only the solution
\begin{eqnarray}
\label{eqn:HauptGleichungQ2}
q2 &=& \left(-1+3 x-x^2-x^3-x^4+x^5-3 x^6+x^7+x^8+x^9+ \right.\\
\nonumber
&& \left. \surd 
\left(1-6 x+11 x^2-4 x^3-3 x^4-6 x^5+15 x^6-16 x^7+x^8+2 x^9+9 x^{10}-
\right. \right. \\
\nonumber
&& \left. \left. 8 x^{11}+7
x^{12}+6 x^{13}+5 x^{14}+3 x^{16}+2 x^{17}+x^{18}\right)\right)/\left(x^2-2 x^3-x^4+x^6+3 x^8+2 x^9+x^{10}\right)
\end{eqnarray}
is possible, since its Taylor expansion is
\begin{eqnarray*}
q2 &=& 2 x^5+6 x^6+16 x^7+40 x^8+96 x^9+228 x^{10}+532 x^{11}+1230 x^{12}+2826 x^{13}+6464 x^{14}+14742 x^{15}+ \\
&&33546 x^{16}+76216 x^{17}+172968
x^{18}+392228 x^{19}+888932 x^{20}+O(x)^{21}.
\end{eqnarray*}
These values agree with those from the following table that
were obtained by our dynamic programming implementation.
\medskip

\noindent
\begin{tabular}{|l|rrrrrrrrrrrrrrrrrr|}
\hline
$n$&0&1&2&3&4&5&6&7&8&9&10&11&12&13&14&15&16&17\\
\hline
$Q_n$&0&0&0&0&0&2&6&16&40&96&228&532&1230&2826&6464&14742&33546&76216\\
\hline
\end{tabular}
\medskip

\subsection{Flajolet-Odlyzko theorem}

Having determined the formulas for $z1$ [resp. $q2$] in
equation (\ref{eqn:HauptGleichungZ1})
[resp. equation (\ref{eqn:HauptGleichungQ2}], we use the following
theorem of Flajolet and Odlyzko \cite{FlaOdl90} to determine the asymptotic
coefficients of $Z_n$  [resp. $Q_n$] in the Taylor expansion of
the generating function $z1$ [resp. $q2$]. 

\begin{figure}[ht]
\begin{center}
\includegraphics[width=0.55\linewidth]{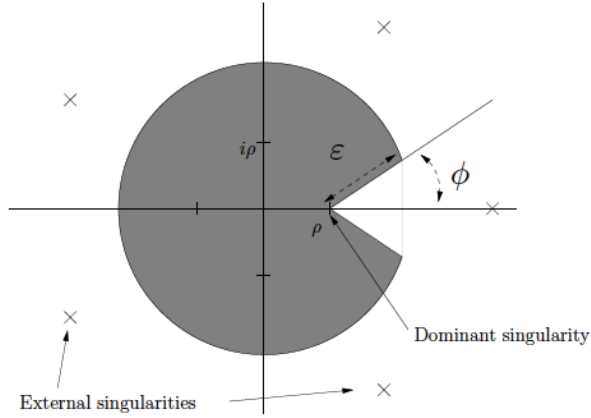}
\end{center}
\caption{The shaded region $\triangle$ where, except at $x=\rho$, the generating
 function $f(x)$ must be analytic. Here $\rho=1$.
Figure taken from Lorenz et al. \cite{Lorenz.jcb08}.}
\label{fig:FlajoletTriangle}
\end{figure}

Following standard convention, let $[x^n]f(x)$ denote the $n$th
coefficient in the Taylor expansion of $f(x)$.
The following theorem is stated as Corollary 2, part (i) 
of \cite{FlaOdl90} on page 224.
\begin{theorem}[Flajolet  and Odlyzko]
\label{thm:flajolet}
Assume that $f(x)$ has a singularity at $x=\rho>0$, is analytic in
the rest of the region $\triangle \backslash {1}$, depicted in
Figure \ref{fig:FlajoletTriangle},
and that as $x \rightarrow\rho$ in $\triangle$,
\begin{equation}
\label{alg-sing}
f(x) \sim K(1-x/\rho)^{\alpha}.
\end{equation}
Then, as $n \rightarrow \infty$, if $\alpha \notin {0, 1, 2, ...}$,
\begin{eqnarray*} 
f_n = [x^n]f(x) \sim \frac{K}{\Gamma(-\alpha)} \cdot n^{-\alpha-1}\rho^{-n}
\end{eqnarray*}
where $\Gamma$ denotes the Gamma function.
\end{theorem}

In Section \ref{section:asymptoticValueQ2}, we determine the asymptotic
value of $[x^n]q2 = Q_n$, and in the following
Section \ref{section:asymptoticValueZ1}, we determine the asymptotic
value of $[x^n]z1 = Z_n$. The ratio of these values then yields the
asymptotic expected number of neighbors, or {\em network connectivity} for 
the homopolymer model of RNA.

\subsection{Asymptotic number of neighbors}
\label{section:asymptoticValueQ2}

Let $P$ denote the polynomial under the radical of 
equation~(\ref{eqn:HauptGleichungQ2}), i.e.
\begin{eqnarray}
P &=& 1-6 x+11 x^2-4 x^3-3 x^4-6 x^5+15 x^6-16 x^7+x^8+2 x^9+ \\
&&9 x^{10}-8 x^{11}+7 x^{12}+6 x^{13}+ 5 x^{14}+3 x^{16}+2 x^{17}+x^{18}.
\end{eqnarray}
There are 4 real roots and 14 imaginary roots of $P$; however, the
root of the smallest modulus (absolute value) is the real root 
\begin{eqnarray}
\label{eqn:rhoForQ}
\rho =  0.436911127214519 \approx 0.436911.
\end{eqnarray}
Now $q2$ can be expressed in the form
$q2=G+H$, where $G=\frac{Gnum}{Gdenom}$ and
$H=\frac{\sqrt{P}}{Gdenom}$, and where
\begin{eqnarray}
\label{eqn:GnumGdenomForQ}
Gnum &=& \left(-1+3 x-x^2-x^3-x^4+x^5-3 x^6+x^7+x^8+x^9\right)\\
Gdenom &=& x^2 \left(1-2 x-x^2+x^4+3 x^6+2 x^7+x^8\right).
\nonumber
\end{eqnarray}
One notes that $\rho$ is a root of both $Gnum$ and $Gdenom$, so
their ratio is well-defined; as well, clearly
$0$ is a root of $Gdenom$.  It follows that both $0$ and $\rho$
are singularities of the function $q2$ (recall that in complex analysis,
the square root of $0$ is a singularity).

For asymptotics, the term
$\frac{Gnum}{Gdenom}$ can be neglected, since 
$\lim_{x \rightarrow \rho} \frac{Gnum}{Gdenom} =
-2.963617606602476$; thus it follows that
$[x^n]q2 = [x^n]\frac{\sqrt{P}}{Gdenom}$. However, the
Flajolet-Odlyzko theorem cannot be applied 
to the function $f=\frac{\sqrt{P}}{Gdenom}$, since $\rho$ is not
the dominant singularity, as $0 < \rho$.
To address this issue, we define
$Gnum1=Gnum$ and $Gdenom1 =Gdenom / x^2$, hence
\begin{eqnarray}
\label{eqn:GnumGdenomForQ1}
Gnum1 &=& \left(-1+3 x-x^2-x^3-x^4+x^5-3 x^6+x^7+x^8+x^9\right)\\
Gdenom1 &=& \left(1-2 x-x^2+x^4+3 x^6+2 x^7+x^8\right).
\nonumber
\end{eqnarray}
Now we can apply Theorem~\ref{thm:flajolet} to the function 
$f1 = x^2 \cdot f = \frac{\sqrt{P}}{Gdenom1}$, for which $\rho$ is
the dominant singularity. 
First, we factor $(1-x/\rho)$ out from $P$. 
\begin{eqnarray}
\label{eqn:PdividedByOneMinusXoverRho}
\frac{P}{1-x/\rho} &=&
1-3.71121 x+2.50581 x^2+1.73529 x^3+0.971726 x^4-3.77592 x^5 +6.3577 x^6-\\
&&
1.44854 x^7-2.3154 x^8-3.29948 x^9+1.44816 x^{10}-4.68545 x^{11}- 
3.72404 x^{12}-\\
&&
2.52356 x^{13}-0.775919 x^{14}-1.77592 x^{15}-1.06471 x^{16}-0.436911 x^{17}.
\end{eqnarray}
Second, we factor $(1-x/\rho)$ out from $Gdenom1$. 
\begin{eqnarray}
\label{eqn:Gdenom1dividedByOneMinusXoverRho}
\frac{Gdenom1}{1-x/\rho} &=&
1+0.288795 x-0.339007 x^2-0.775919 x^3-0.775919 x^4-1.77592 x^5-\\
&&
1.06471 x^6-0.436911 x^7.
\nonumber
\end{eqnarray}
It follows from (\ref{eqn:PdividedByOneMinusXoverRho}) and
(\ref{eqn:Gdenom1dividedByOneMinusXoverRho}) that
\begin{eqnarray}
\label{eqn:KforQ}
\frac{\sqrt{P}}{Gdenom1} &=&
\frac{(1-x/\rho)^{-1/2} \cdot \sqrt{P/(1-x/\rho)}} {Gdenom1/(1-x/\rho)} \\
\frac{\sqrt{P/(1-x/\rho)}} {Gdenom1/(1-x/\rho)}\left( \rho \right) &=&
0.11422693623949792 . 
\end{eqnarray}
and so 
$\frac{\sqrt{P}}{Gdenom1} = 0.11422693623949792 \cdot (1-x/\rho)^{-1/2}$.
If we define $K = 0.11422693623949792$, then it follows that
$f1 = \frac{\sqrt{P}}{Gdenom1} \sim K(1-x/\rho)^{-1/2}$, and by 
applying Theorem~\ref{thm:flajolet} for $\alpha=-1/2$, we have the
following asymptotic result.
\begin{lemma}
\label{thm:asymptoticQ} 
The asymptotic value of the $n$th coefficient
$[ x^n ]\left(x^2 q2 \right)$  in the Taylor expansion of
$x^2 \cdot \sum_{n=0}^{\infty} Q_n x^n$ is
$\frac{0.06444564758689844 \cdot 2.2887949921884863^n} {\sqrt{n}}$.
\end{lemma}
\smallskip

\noindent
{\sc Proof.} We have
\begin{eqnarray}
\nonumber
[ x^n ]\left(x^2 q2 \right) &=&
[x^n]\left(f1\right) \sim \frac{K}{\Gamma(-\alpha)} \cdot
n^{-\alpha-1} \cdot \rho^{-n}\\
 &=&
\nonumber
\frac{0.11422693623949792}{\Gamma(1/2)} \cdot
n^{-1/2} \cdot 0.436911127214519^{-n} \\
 &=&
\frac{0.06444564758689844 \cdot 2.2887949921884863^n}
{\sqrt{n}} \hfill \Box
\end{eqnarray}

\subsection{Asymptotic number of structures}
\label{section:asymptoticValueZ1}

We now proceed similarly to determine the asymptotic value of
the Taylor coefficients of $x^2 \cdot \sum_{n=0}^{\infty} Z_n x^n$.
Now let $P$ denote the polynomial under the radical of 
equation~(\ref{eqn:HauptGleichungZ1}), i.e.
\begin{eqnarray}
P &=& 1-2 x-x^2+x^4+3 x^6+2 x^7+x^8.
\end{eqnarray}
There are 2 real roots and 6 imaginary roots of $P$; however, the
root of the smallest modulus (absolute value) is the real root 
\begin{eqnarray}
\label{eqn:rhoForZ}
\rho =  0.436911127214519 \approx 0.436911,
\end{eqnarray}
identical the the value from equation~(\ref{eqn:rhoForQ}).
Then $z1$ can be expressed in the form
$z1=G+H$, where $G=\frac{Gnum}{Gdenom}$ and
$H=\frac{\sqrt{P}}{Gdenom}$, and where
\begin{eqnarray}
\label{eqn:GnumGdenomForZ}
Gnum &=& \left( 1 - x + x^2 + x^3 + x^4 \right)\\
Gdenom &=& 2 x^2.
\nonumber
\end{eqnarray}
Note that $\rho$ is not a root of either $Gnum$ or $Gdenom$, so
their ratio is well-defined; as well, clearly
$0$ is a root of $Gdenom$.  It follows that both $0$ and $\rho$
are singularities of the function $z1$ (recall that in complex analysis,
the square root of $0$ is a singularity).

For asymptotics, the term
$\frac{Gnum}{Gdenom}$ can be neglected, since 
$\lim_{x \rightarrow \rho} \frac{Gnum}{Gdenom} =
2.2887949921884205$; thus it follows that
$[x^n]z1 = [x^n]\frac{\sqrt{P}}{Gdenom}$. However, the
Flajolet-Odlyzko theorem cannot be applied 
to the function $f=\frac{\sqrt{P}}{Gdenom}$, since $0$, rather than
$\rho$, is the dominant singularity.
To address this issue, we define
$Gnum1=Gnum$ and $Gdenom1 =Gdenom / x^2$, hence
\begin{eqnarray}
\label{eqn:GnumGdenomForZ1}
Gnum1 &=& \left( 1 - x + x^2 + x^3 + x^4 \right)\\
Gdenom1 &=& 2.
\nonumber
\end{eqnarray}
Now we can apply Theorem~\ref{thm:flajolet} to the function 
$f1 = x^2 \cdot f = \frac{\sqrt{P}}{Gdenom1}$, for which $\rho$ is
the dominant singularity. 
First, we factor $(1-x/\rho)$ out from $P$. 
\begin{eqnarray}
\label{eqn:PdividedByOneMinusXoverRhoForZ}
\frac{P}{1-x/\rho} &=&
1 + 0.288795 x - 0.339007 x^2 - 0.775919 x^3 - 0.775919 x^4 - 
  1.77592 x^5 - 1.06471 x^6 - 0.436911 x^7.
\end{eqnarray}
It follows from (\ref{eqn:PdividedByOneMinusXoverRhoForZ}) that
\begin{eqnarray}
\label{eqn:KforZ}
\frac{\sqrt{P}}{Gdenom1} &=&
\frac{(1-x/\rho)^{1/2} \cdot \sqrt{P/(1-x/\rho)}} {Gdenom1} \\
\frac{\sqrt{P/(1-x/\rho)}} {Gdenom1}\left( \rho \right) &=&
\frac{\sqrt{P/(1-x/\rho)}} {2}\left( \rho \right) =
0.4825630725501931
\end{eqnarray}
and so 
$\frac{\sqrt{P}}{2} = 0.4825630725501931 \cdot (1-x/\rho)^{1/2}$.
If we define $K = 0.4825630725501931$, then it follows that
$f1 = \frac{\sqrt{P}}{2} \sim K(1-x/\rho)^{1/2}$, and by 
applying Theorem~\ref{thm:flajolet} for $\alpha=+1/2$, we have the
following asymptotic result.
\begin{lemma}
\label{thm:asymptoticZ} 
The asymptotic value of the $n$th coefficient
$[ x^n ]\left(x^2 z1 \right)$  in the Taylor expansion of
$x^2 \cdot \sum_{n=0}^{\infty} Z_n x^n$ is
$\frac{0.13612852946880957 \cdot 2.2887949921884863^n}{n^{3/2}}$.
\end{lemma}
\smallskip

\noindent
{\sc Proof.} We have
\begin{eqnarray}
\nonumber
[ x^n ]\left(x^2 z1 \right) &=&
[x^n]\left(f1\right) \sim \frac{K}{\Gamma(-\alpha)} \cdot
n^{-\alpha-1} \cdot \rho^{-n}\\
 &=&
\nonumber
\frac{0.13612852946880957}{\Gamma(-1/2)} \cdot n^{-3/2}
\cdot 2.2887949921884863^n \\
 &=&
\frac{0.13612852946880957 \cdot 2.2887949921884863^n}{n^{3/2}}.
\end{eqnarray}
$\Box$
We have now established the asymptotic {\em expected} network connectivity.
\begin{theorem}
\label{thm:asymptoticQoverZ}
The asymptotic value of the expected number of neighbors
is $0.4734176431521986 \cdot n$; i.e. the asymptotic value of
$\sum_{n=0}^{\infty} \frac{Q_n}{n Z_n} \cdot x^n$ is
$0.4734176431521986$.
\end{theorem}
\smallskip

\noindent
{\sc Proof.} By 
Lemmas~\ref{thm:asymptoticQ} and \ref{thm:asymptoticZ},
we have
\begin{eqnarray*}
\frac{[x^n]q2}{[x^n]z1} &=&
\frac{[x^n](x^2 \cdot q2)}{[x^n](x^2 \cdot z1)}\\
&=&
\frac{(0.06444564758689844 \cdot 2.2887949921884863^n \cdot n^{-1/2}}
{0.13612852946880957 \cdot 2.2887949921884863^n \cdot n^{-3/2}}\\
&=& 0.4734176431521986 \cdot n.
\end{eqnarray*}

\section{Acknowledgements}

This research was funded by the 
National Science Foundation grant DBI-1262439.
Akademischer Austauschdienst.  Any opinions, findings,
and conclusions or recommendations expressed in this material are
those of the authors and do not necessarily reflect the views of the
National Science Foundation.

\bibliographystyle{plain}
\bibliography{biblio}
\end{document}